# Anomalous dependence of thermoelectric parameters on carrier concentration and electronic structure in Mn-substituted Fe$_2$CrAl Heusler alloy

Kavita Yadav[1], Saurabh Singh[2], Omprakash Muthuswamy[2], Tsunehiro Takeuchi[2], and K. Mukherjee[1*]

[1]School of Basic Sciences, Indian Institute of Technology, Mandi, Himachal Pradesh-175005, India

[2]Research Centre for Smart Energy Technology, Toyota Technological Institute, Nagoya, 468-8511, Japan

*Corresponding Author: K. Mukherjee

    School of Basic Sciences,

    Indian Institute of Technology Mandi,

    Mandi 175005, Himachal Pradesh, India

    Tel: 0091 1905 267043

    E-mail: kaustav@iitmandi.ac.in

# Anomalous dependence of thermoelectric parameters on carrier concentration and electronic structure in Mn-substituted $Fe_2CrAl$ Heusler alloy

**Abstract**

We investigate the high temperature thermoelectric properties of Heusler alloys $Fe_{2-x}Mn_xCrAl$ ($0 \leq x \leq 1$). Substitution of 12.5% Mn at Fe-site (x = 0.25) causes a significant increase in high temperature resistivity ($\rho$) and an enhancement in the Seebeck coefficient (*S*), as compared to the parent alloy. However, as the concentration of Mn is increased above 0.25, a systematic decrement in the magnitude of both parameters is noted. These observations have been ascribed (from theoretical analysis) to a change in band gap and electronic structure of $Fe_2CrAl$ with Mn-substitution. Due to absence of mass fluctuations and lattice strain, no significant change in thermal conductivity ($\kappa_l$) is seen across this series of Heusler alloys. Additionally, *S* drastically changes its magnitude along with a crossover from negative to positive above 900 K, which has been ascribed to the dominance of holes over electrons in high temperature regime. In this series of alloys, *S* and $\rho$ shows a strong dependence on the carrier concentration and strength of *d-d* hybridization between Fe/Mn and Cr atoms.

## 1. Introduction

Heusler alloys provide a platform where novel physical phenomenon like thermoelectric effect, magneto-optical effect, magnetocaloric effect, unconventional superconductivity, half-metallicity, magnetic shape memory effect, barocaloric effects, magnetoresistance, anomalous Hall Effect and large exchange bias effect can be studied [1-12]. Thermoelectric properties of half-Heusler alloys have also been studied extensively. However, the possibility of using full Heusler alloys as potential thermoelectric materials has received less attention. However, full Heusler alloys show metallic type conduction, which limits their use as thermoelectric material. But recent theoretical studies on them show that $Co_2$-based and $Fe_2$-based full Heusler alloys exhibit half-metallic ground state with ~ 100 % spin polarization at Fermi level; where, majority spin channel show metallic and minority spin channels show insulating behaviour [13-16]. As per experimental literature reports, the presence of semiconducting ground state in full Heusler alloys are very rare [17].

Among full Heusler alloys, $Fe_2VAl$ is one of the most interesting and important examples which show semi-metallic ground state and interesting thermoelectric properties [18-20]. At the stoichiometric composition, Seebeck coefficient ($S$) is found to be ~ -20 µV/K near 300 K. However, recent reports [20] show that doping of quaternary elements such as Si leads to significant decrease in the low temperature resistivity and enhancement of the $S$. As $S$ is very sensitive to the electron density at the Fermi level, the excess of valence electrons in $p$-states of Si atom can be responsible for the significant decrease in $\rho$ and drastic change in $S$. From the transport studies of Ge-substituted $Fe_2VAl$ alloys, it was concluded that doping of heavier element does affect the transport properties [18, 21]. From the perspective of thermoelectric effect, one of the least studied alloys of this category is $Fe_2CrAl$. N. Arikan *et al.* [22] have predicted the half metallic behaviour in this alloy, where spin down channel exhibit a band gap of ~ 0.2912 eV. However, there are no experimental reports on thermal transport properties of this alloy in the high temperature regime (300-1000 K). S. Fujii *et al.*, [23] have performed first principle calculations on similar alloys, $Fe_2MnZ$ (Z = Al, Si, P), and found that the Z atom influences the electronic behaviour of the system through hybridization between $p$-states of Z element with $d$-states of Fe and Mn. This suggests that hybridization plays an important role in the formation of band gap in these systems. The most pertinent questions related to these types of Heusler alloys are whether the hybridization of the XY- $d$-states and Z- $p$-states alone can account for the observed changes in the transport behaviour or the size of the atoms/density of the valence electrons provided by the X/Y/Z atoms can influence the transport behaviour of such systems. Hence, in order to shed some light on the above-mentioned issues, in the present study, we have probed the evolution of the high temperature transport properties with the substitution of quaternary Mn element at Fe-site of Fe$_2$CrAl.

From our experimental and theoretical results, it is revealed that the magnitude of $S$ and $\rho$ for Fe$_{1.75}$Mn$_{0.25}$CrAl is increased as compared to the Fe$_2$CrAl. Even though, holes are introduced in the parent alloy, minor change in electronic structure around Fermi level ($E_F$) is noted, implying that, inspite of hole doping, electrons play a significant role. Increment in $S$ can be ascribed to the decrease in charge carrier concentration, which is due to the charge compensation effect. However, in Fe$_{1.5}$Mn$_{0.5}$CrAl, Fe$_{1.25}$Mn$_{0.75}$CrAl and FeMnCrAl, $S$ and $\rho$ decreases due to the increment in carrier density with Mn-substitution and the absence of band gap in spin down channel. Our study reveals that, in addition to contribution from carrier concentration, electronic structure of the Fe$_{2-x}$Mn$_x$CrAl ($0 \leq x \leq 1$) plays a crucial role in the observed anomalous transport behaviour.

## 2. Experimental and computational details:

The Heusler alloys Fe$_{2-x}$Mn$_x$CrAl (x = 0, 0.25, 0.5, 0.75 and 1) are prepared by arc melting the stoichiometric ratio of the high purity of the constituent elements (>99.9%) in an atmosphere of argon. The resultant ingots are sealed in evacuated quartz tubes and subjected to 900 °C for 1 week, followed by quenching. Structural characterizations of these alloys have already been reported in Ref. [24]. All these alloys crystallize in cubic L2$_1$ (*Fm-3m* space group). Through Le-bail fit of room temperature synchrotron XRD data (Fig. 1 of Ref. 24), we have confirmed the existence of anti-site disorder in the parent alloy. It was also found that anti-site disorder increases with the Mn concentration. The morphological and elemental analysis of the alloys are done by scanning electron microscope (SEM) using JEOL JSM- 6330F and energy dispersive x-ray spectroscopy (EDX) using a JED-2140GS with an accelerating voltage of 20 kV, respectively. The compositional analysis is carried out by electron probe micro analysis (EPMA) by JEOL JXL-8230. Seebeck and electrical resistivity are measured by the steady state method and four probe technique [25]. Thermal conductivity ($\kappa$) measurements are done by using laser flash analysis (NETZSCH LFA 457). Room temperature hall coefficient measurements are carried in Physical property measurement system (PPMS), Quantum Design, U.S.A. Density of respective alloys is determined by using the Archimedes principle. To get better insight about the transport properties of this series of alloys, we have also performed the density functional theory (DFT) calculations with generalized gradient approximation (GGA). It is in the form of Perdue–Burke–Ernzerhof (PBE) for the exchange-correlation energy functional, as implemented in the WIEN2K code. For all the calculations of electronic density of states, the highly dense k-mesh size of 20×20×20 is used. A conventional unit cell and supercell (2X1X1) are used for the calculations of $Fe_2CrAl$ and Mn doped alloys, respectively. The theoretical lattice parameters obtained from DFT calculations agree with the experimental lattice parameters (given in table S7 of the supplementary information). The DFT calculations have been performed using the experimentally obtained lattice parameters. These parameters are determined using Le-bail fit of room temperature synchrotron XRD pattern [24], which gives us an idea about the atomic distribution in the unit cell. The Fe/Mn atoms are placed on 8a (1/4, 1/4, 1/4), and the Cr and Al atoms are placed on 4a (0, 0, 0) and 4b (1/2, 1/2, 1/2) Wyckoff positions, respectively. A high dense k-mesh of 50 X 50 X 50 points is taken for the self-consistency field (SCF) as well as for the calculation of thermoelectric properties. The energy convergence criteria for the SCF calculations are set to the $10^{-6}$ Rydberg. For the thermoelectric parameters, transport coefficients calculations are performed using Semi-classical Boltzmann transport theory implemented in the BoltzTraP code [26, 27]. These codes provide the parameters within the constant relaxation time approximation.

## 3. Results and Discussions-
### 3.1. Surface Morphological analysis

As mentioned in the previous section, from our recent experimental report, it was concluded that the degree of anti-site disorder between Fe and Al atoms increases with increment in Mn concentration at Fe-site [24]. These disorders within the lattice can influence the thermoelectric behaviour of intermetallics [1-3]. To ascertain the quality of our alloys, surface (SEM) and compositional morphology (EDX) studies were carried out.

Figure 1(a)-(e) shows the SEM and EDX of $Fe_{2-x}Mn_xCrAl$ (x = 0, 0.25, 0.5, 0.75 and 1). It indicates that all the alloys have smooth surface with homogenous distribution of elements within the alloy. From the EPMA analysis, the average chemical composition are determined as $Fe_{2.011\pm0.006}Cr_{1.094\pm0.004}Al_{0.8945\pm0.05}$, $Fe_{1.782\pm0.004}Mn_{0.245\pm0.001}Cr_{1.082\pm0.03}Al_{0.8916\pm0.02}$, $Fe_{1.543\pm0.003}Mn_{0.504\pm0.003}Cr_{1.003\pm0.003}Al_{0.949\pm0.007}$, $Fe_{1.289\pm0.012}Mn_{0.710\pm0.008}Cr_{1.079\pm0.013}Al_{0.921\pm0.017}$, and $Fe_{1.094\pm0.004}Mn_{0.891\pm0.009}Cr_{1.113\pm0.009}Al_{0.902\pm0.05}$, corresponding to the nominal composition of $Fe_2CrAl$, $Fe_{1.75}Mn_{0.25}CrAl$, $Fe_{1.5}Mn_{0.5}CrAl$, $Fe_{1.25}Mn_{0.75}CrAl$ and FeMnCrAl, respectively. The assigned chemical composition of each alloy is summarized in the supplementary material (Tables S1-S5). The density values obtained from Archimedes method are 6.435,

6.373, 6.377, 6.315 and 6.289 g/cc, for $Fe_2CrAl$, $Fe_{1.75}Mn_{0.25}CrAl$, $Fe_{1.5}Mn_{0.5}CrAl$, $Fe_{1.25}Mn_{0.75}CrAl$ and FeMnCrAl, respectively. The theoretical density is mentioned in the table S6 of supplementary material. According to literature reports good quality thermoelectric materials should have a very high measured density [28]. In our case, it is found that the measured density is greater than the 97% of theoretical density implying high density of all the alloys.

### 3.2. Electrical resistivity-

To study the high temperature thermoelectric properties of $Fe_{2-x}Mn_xCrAl$ (0≤x≤1) alloys, $S$, $\rho$ and $\kappa$ measurements are carried out. This section gives the details of electrical resistivity of this series of alloys. Figure 2 represents the temperature response of the $\rho$ in the temperature range 300-950 K. $Fe_2CrAl$ exhibits a semiconducting-like behaviour with negative temperature coefficient of electrical resistivity reaching ~ 0.20 mΩ-cm at 1000 K, which is consistent with band structure calculation, discussed later in Section 3.5. By the partial substitution of Mn at Fe site, the magnitude of $\rho_{300\,K}$ increases significantly and reaches 0.33 mΩ-cm for $Fe_{1.75}Mn_{0.25}CrAl$. This can be attributed to the charge compensation effect arising due to a decrement in electron concentration on substitution of $p$-type Mn.

Also, $Fe_{1.5}Mn_{0.5}CrAl$, $Fe_{1.25}Mn_{0.75}CrAl$ and FeMnCrAl exhibit semiconducting-like behaviour. However, from band structure calculations, the latter three alloys are predicted to exhibit metallic behaviour (discussed in later section), which is contradictory to the observed experimental behaviour. As mentioned before, there is a presence of anti-site disorder between Fe and Al in these alloys which leads to the observation of the semiconducting behaviour of $\rho$. This type of disorder can alter the band structure and affect the transport behaviour of the Mn doped alloys. Similar type of features was noted in $Fe_2VAl$ and $Fe_2TiSn$ [18-20, 30], where despite of having metallic constituents, they exhibit semiconducting behaviour.

On increment of Mn content beyond 12.5%, a systematic decrement in value of $\rho_{300\,K}$ is noted (inset of Fig. 2). The value of $\rho$ drops from $\rho_{300K}$ ~ 0.33 mΩ-cm in $Fe_{1.75}Mn_{0.25}CrAl$ to $\rho_{300K}$ ~ 0.29 mΩ-cm in FeMnCrAl. This trend can be ascribed to reduction in charge compensation effect due to the presence of anti-site disorder in higher Mn- substituted alloys. The observed trend of the magnitude of $\rho$ in this series of alloy is consistent with literature report on a similar doped metallic system [31, 32]. Additionally, for all Mn compositions, the $\rho$ curves exhibit similar temperature dependence as that of parent alloy till 1000 K. Interestingly, in all these alloys, a broad hump-like feature is noted near 600 K. This feature can be due to bipolar effect. This effect is also observed in temperature response of *S* and is discussed in detail in later section. However, we have also noted an additional increment in magnitude of *ρ* above 800 K. It can be due to presence of n to p transition, which is visible in temperature response of Seebeck coefficient (discussed in later section). With further increase in temperature, no overlap between respective *ρ* curves is observed till 950 K. This is contrast to that reported in Ta-doped Fe$_2$VAl [33] and W-doped Fe$_2$VAl [34] systems, where these curves coincide at higher temperatures (~ 900 K). In these later systems, when all the carriers are excited across the gap, the value of *ρ* becomes constant for all doping concentrations.

### 3.3. Seebeck coefficient

The '*S*' measured as a function of temperature in the range 300-950 K for Fe$_{2-x}$Mn$_x$CrAl (0≤x≤1) alloys are shown in Figure 3. At 300 K, all the alloys exhibit negative *S* which suggests that electrons are the dominant charge carriers. For the parent alloy Fe$_2$CrAl, the value of *S* is found to be ~ -6.68 µV/K near 300 K. The observed magnitude and sign of the *S* is in agreement with the earlier reported value [31]. Such a moderate value at room temperature is a characteristic of metallic state and indicates the possibility of half-metallic ground state. This observed value of *S* is consistent with the value obtained by DFT calculations [section 3.5]. For Fe$_{1.75}$Mn$_{0.25}$CrAl, an enhancement in the absolute value of *S* is noted (~ -9.95 µV/K), whereas, for alloys with higher Mn concentration, a systematic decrement in the magnitude is observed (inset of Fig. 3). This sudden increment in magnitude of *S* in Fe$_{1.75}$Mn$_{0.25}$CrAl can arise as a result of charge compensation effect, while in higher Mn concentration this effect diminishes due to the presence of anti-site disorder. This behaviour is contrast to that noted in Mn$_2$VAl alloys [35]. In these alloys, a systematic increment in magnitude of *S* with increment in anti-site disorder was observed.

Interestingly, the variation in the magnitude of *S* with Mn concentration (at 300 K) is noted to be non-monotonic. In order to determine the carrier concentration (*n*), Hall resistance (*R*$_H$) is measured at 300 K. *R*$_H$ is expressed as

$$R_\mathrm{H} = \frac{-1}{n\,e} \ldots\ldots\ldots\ldots\ldots\ldots (1)$$

The observed Hall coefficient is negative, indicating that majority charge carriers are electrons. The obtained value of *n* corresponding to each alloy is given in table 1. For Fe$_2$CrAl, *n* is calculated ~ 2.16 X 10$^{21}$/cm$^3$ which is higher than that noted in Fe$_2$VAl (10$^{20}$/cm$^3$). It is comparable with that noted in metallic and highly degenerate semiconductors

[36, 37] However, in $Fe_{1.75}Mn_{0.25}CrAl$ the value of $n \sim 9.8 \times 10^{20}/cm^3$ which is one order smaller than that obtained in parent alloy. With further increment in Mn-concentration, a monotonic increment in magnitude of $n$ is noted, which is in accordance with the observed temperature dependent behaviour of $S$. It implies that as $n$ increases, there is decrease in magnitude of $S$. It is also interest to note that inspite of difference between magnitudes of $n$ in $Fe_2CrAl$ and $Fe_{1.5}Mn_{0.5}CrAl$, temperature dependent $S$ curve for both alloys overlaps. This indicates that, other than $n$, there are other factors which play an important role in the observed behaviour of $S$. Similar type of behaviour with hole substitution is also reported in $Fe_2Ti_{1-x}Cr_xAl$ system, where small Ti substitution at Cr site increases the magnitude of $S$, and further increase in substitutions make it *p*-type due to the dominance of hole charge carriers [31, 32]. In the present study, such type of unusual feature is also observed with Mn-substitution, which can be attributed to the modification in the electronic structure.

All the alloys show similar temperature dependent behaviour. For each composition, with increment in temperature, initially the value of $S$ increases and attains a maximum near 550 K. The maximum value of $S$ is found $\sim -12.17$ µV/K for $Fe_{1.75}Mn_{0.25}CrAl$. After 550 K, a downturn in $S$ with temperature is observed, and $|S|$ start decreasing with further increment in temperature. Such type of variation in the $S$ can be due to the presence of bipolar effect at high temperature (above 500 K) in all alloys. Both majority (electrons) and minority (holes) carriers are in significant number and contribute to $S$, resulting in decrement in its magnitude due to offsetting of contribution of majority carriers by minority carriers. This phenomenon occurs at high temperature when the electrons are excited across the band gap producing minority charge carriers (e.g. holes in present case) in addition to majority charge carriers (e.g. the electrons in present case). Such a behaviour usually observed in alloys having small band gap. Above 750 K, where holes dominate, the magnitude of $S$ systematically decreases with increase in the concentration of Mn. Interestingly, a change in the sign of $S$ from negative to positive is noted above 900 K in all the alloys. The observed sign change indicates the dominance of hole as charge carriers above 900 K. Similar features has been reported in other Heusler alloys [38].

### 3.4. Thermal conductivity

To further evaluate the thermoelectric performance of this series, $\kappa$ is measured in the temperature range of 300-950 K (Fig. 4 (a)). In general, the total thermal conductivity ($\kappa_{total}$) of metals and semimetals is defined as by a sum of electronic ($\kappa_e$) and lattice ($\kappa_L$) contributions. For better understanding about the contribution of electrons in $\kappa_{total}$, a modified equation for $\kappa_e$ can be used, which is given as

$$k_e = \left[1.5 + exp^{-\left[\frac{|S|}{116}\right]}\right]\sigma T \ldots\ldots\ldots (2)$$

where $S$, $\sigma$, and $T$ represent the Seebeck coefficient, electrical conductivity, and absolute temperature, respectively [28]. Here, the lattice thermal conductivity ($\kappa_L$) can be evaluated by subtracting $\kappa_e$ from the observed $\kappa$. The temperature response of $\kappa_e$ and $\kappa_L$ for alloys is shown in Fig 4 (b) and inset of Fig. 4(a), respectively. In case of all alloys, it can be observed that $\kappa_e$ increases with increment in temperature and with increment in Mn substitution, negligible change in magnitude of $\kappa_e$ is noted. Interestingly, in this series of alloys, transfer of thermal energy is mainly due to electrons rather than lattice vibrations i.e., $\kappa_e > \kappa_l$. This behaviour is similar to that noted in metals and alloys.

At 300 K, in case of Fe$_2$CrAl, the value of $\kappa_L$ is ~ 5.05 W m$^{-1}$ K$^{-1}$, which is much smaller than those found in Fe$_2$VAl (~ 28 W m$^{-1}$ K$^{-1}$) and Fe$_2$VGa (~ 17 W m$^{-1}$ K$^{-1}$) [18, 39]. This reduction can be attributed to anti-site defect arising from disorder between Fe and Al. Above 600 K, parabolic behaviour in $\kappa_L$ is seen, which is consistent with the existence of both holes and electrons. This feature is in analogy with the presence of bipolar effect as noted in temperature dependent $\rho$ and S. Interestingly, a hump is noted near the $n$ to $p$ type transition temperature (800 K). Across the series, at 300 K, no significant change in value of $\kappa_L$ is noted. The factors which influence the lattice thermal conductivity are mass fluctuations and lattice strain, caused by introduction of the additional element [33, 34]. In the present case, Fe and Mn have comparable atomic weight, thus mass fluctuation effect can be considered negligible in all these alloys. Crystal lattice strain arises due to volume fluctuations and anti-site defects which is induced due to elemental substitution. Due to lattice strain, broadening and reduction in intensity of XRD peaks will be observed. However, for these alloys, no significant broadening in the XRD peaks and reduction in intensity of (200) peak is noted with increment in Mn substitution [24]. It implies that change in lattice thermal conductivity due to crystal lattice strain with Mn-substitution is negligible in the present case. However, the magnitude of the observed lattice thermal conductivity is larger as compared to highly efficient thermoelectric materials like Bi$_2$Te$_3$, Sb$_2$Te$_3$ [40]. The larger values of $\kappa_e$ and $\kappa_L$ can affect the magnitude of $S$, which leads to decrease in efficiency of thermoelectric materials. In order to reduce the $\kappa_L$, one can substitute heavier and larger radii atoms such as W, Nb etc. at Fe-site. These heavier and larger radii dopants can increase the phonon scattering. Whereas $\kappa_e$ can be reduced by manipulating the charge carrier mobility by suitable alloying. For each alloy, the figure of merit (zT) is calculated using the formula [28]

$$zT = \frac{S^2 \sigma T}{K} \quad \ldots\ldots\ldots\ldots\ldots (3)$$

The temperature response of zT for all alloys are shown in inset of Fig. 4 (a). At 950 K, for Fe$_2$CrAl, the value of zT is found to be 0.21 x 10$^{-4}$. With increment in Mn substitution, we have observed a non-monotonous variation in magnitude of zT. The maximum value of zT ~ 0.24 x 10$^{-4}$ is observed in Fe$_{1.75}$Mn$_{0.25}$CrAl. However, this value is very less compared with highly efficient thermoelectric materials [40].

### 3.5. Electronic structure -

From our experimental studies it was observed that a slight replacement of Fe by 12.5 % Mn, increases both $S$ and $\rho$ significantly, whereas on further increment of Mn, a systematic decrease of both parameters is noted. It was also noted that in these series of alloys the magnitude of $S$ is found to be significantly small even though they show a semiconducting behaviour $\rho$ in the measured temperature range. In order to shed some light on these observed unusual behaviours, electronic band structure calculations of this series are carried out. In addition to this, the effect of anti-site disorder between Fe and Al on the electronic band structure of each alloy is discussed and given in supplementary information.

The spin polarized dispersion curves along the direction of high symmetry 1$^{st}$ Brillouin zone (B.Z.) of corresponding alloys are presented in Figures 5-6. In Fe$_2$CrAl, spin-up channel is found to be metallic as 18 highly dispersing bands cross the $E_F$ at 6 different points. In the spin-down channel, one can find a direct band gap ~ 0.318 eV near Γ point. This alloy shows half-metallic ferromagnetic behaviour. This is in analogy with the previous literature report [22]. In Fe$_{1.75}$Mn$_{0.25}$CrAl, this direct band gap decreases to ~ 0.226 eV, whereas in Fe$_{1.5}$Mn$_{0.5}$CrAl, Fe$_{1.25}$Mn$_{0.75}$CrAl and FeMnCrAl, spin-down channel is found to be metallic. This observed feature is ascribed to the shift of bonding energy levels to higher energy states because the number of valence electrons in Fe$_2$CrAl reduces as Mn-substitution is increased. Similar type of behaviour is also noted in Fe$_{1.75}$Cr$_{0.25}$NbAl [41].

The total density of states (TDOS) plot for all the Mn-substituted alloys is shown in Figure 7. In Fe$_2$CrAl, a sharp peak is observed in conduction band above $E_F$. While just below $E_F$, DOS is much lower suggesting that electrons present in conduction band are dominant carriers. This is in analogy with observed negative $S$ in parent alloy. In case of Fe$_{1.75}$Mn$_{0.25}$CrAl, the TDOS decreases at $E_F$, for spin-up channel. The decrement in magnitude of TDOS and a shift towards higher energy near the $E_F$, suggests that hole charge carriers also contribute in the observed transport properties (as shown in right panel of Fig. 7). With further increment in Mn-content, systematic enhancement in the TDOS curvature is found within 1 eV range. TDOS remains almost similar for $Fe_{1.5}Mn_{0.5}CrAl$, $Fe_{1.25}Mn_{0.75}CrAl$ and FeMnCrAl, however, it is decreased by one-third as compared to $Fe_{1.75}Mn_{0.25}CrAl$ at $E_F$, possibly, due to the increment in number of holes with Mn-substitution. Similar trend is also noted for the spin-down channel.

In case of degenerate semiconductors or half metallic systems, linear response theory is generally used to evaluate the electrical conductivity ($1/\rho$). It is mathematically expressed as [42]

$$\sigma(T) = \int_{-\infty}^{\infty} \sigma(\varepsilon, T)(\frac{-\partial f_{FD}(\varepsilon,T)}{\partial \varepsilon})d\varepsilon \ldots\ldots\ldots\ldots (4)$$

where $\sigma(\varepsilon, T)$ and $f_{FD}(\varepsilon, T)$ indicates spectral conductivity and Fermi-Dirac distribution, respectively. Equation 3 suggest that spectral conductivity and energy derivative of the Fermi-Dirac distribution function should be considered to understand the behaviour of $\sigma(T)$. Here, $\frac{-\partial f_{FD}(\varepsilon,T)}{\partial \varepsilon}$ can be calculated at a given temperature. It plays a significant role of a window limiting the energy range of integration [38]. Additionally, $\sigma(\varepsilon, T)$ solely determines the $\sigma(T)$ and can be estimated from the electronic band structure.

From above equation, it can be assumed that in the strong scattering limit, $\sigma(\varepsilon, T)$ is proportional to $N(\varepsilon)$ (density of states at $E_F$). In half-metallic systems or systems exhibiting pseudo-gap, both spin-up and spin-down channel contributes to $N(\varepsilon)$. In $Fe_2CrAl$, there is presence of band gap in spin down channel. Also, there is presence of a large $N(\varepsilon)$ at $E_F$ in spin-up channel (as shown in right panel of Fig. 7 (a)). This leads to lower value of $\rho$ at 300 K. However, in case of $Fe_{1.75}Mn_{0.25}CrAl$, there is a decrement in $N(\varepsilon)$ at $E_F$, which is responsible for the higher value of $\rho$ as compared to parent alloy. Further increment in Mn-content (above 0.25), results in a disappearance of band gap at $E_F$, leading to a contribution of both spin channels to $\sigma(\varepsilon, T)$. It can be noted from Fig. 7 (c)-(e), there is systematic shift in $N(\varepsilon)$ at $E_F$ towards higher energy. This is in analogy with the observed trend in magnitude of $\rho$ at 300 K, in the latter three compositions.

To see the contributions from various atomic states around $E_F$, PDOS plots for all alloys are presented in Figure 8-9. From this figure, it can be said that $s$ and $p$ orbital have minor contribution in the band structure as they are filled. Most of the regions around $E_F$ are occupied by the electrons from $d$-orbital ($e_g$ and $t_{2g}$). In half metallic ferromagnets, $d$-$d$ hybridization is the origin of the band gap in the minority spin channel. The $d$-orbitals are divided into five degenerate states: three-fold degenerate ($t_{2g}$) bonding states and two-fold degenerate ($e_g$) anti-bonding states [44]. The bandgap is formed between the higher $t_{2g}$ bonding hybridised states in the valence band and the lower $e_g$ anti-bonding states in the conduction band of the Fe and Cr $d$-orbitals. The anti-bonding $d$-orbitals of Fe cannot couple with the Cr $d$-orbitals, resulting in a small bandgap near $E_F$. Similar contributions from $t_{2g}$ and $e_g$ states of Fe, Mn and Cr atoms are noted in $Fe_{1.75}Mn_{0.25}CrAl$ (Fig. 8 (c)-(e)). However, with further increment in Mn-concentration, anti-bonding $e_g$-orbitals of Fe/Mn couples with the Cr $d$-orbitals due to formation of hole pockets. This leads to the disappearance of band gap in other three alloys $Fe_{1.5}Mn_{0.5}CrAl$, $Fe_{1.25}Mn_{0.75}CrAl$ and $FeMnCrAl$.

As noticed from TDOS and PDOS of $Fe_2CrAl$ and $Fe_{1.75}Mn_{0.25}CrAl$, both spin-up and spin-down channel contribute towards the $S$. Therefore, in these two alloys, using two current model (Boltztrap code), total contribution to $S$ from both channels can be estimated. The expression of the total $S$ is described by the formula [44, 45]

$$S = \frac{[\sigma(\uparrow)S(\uparrow)+\sigma(\downarrow)S(\downarrow)]}{[\sigma(\uparrow)+\sigma(\downarrow)]} \quad \ldots\ldots\ldots\ldots (5)$$

where $S(\uparrow)$ and $S(\downarrow)$ are the Seebeck coefficients and $\sigma(\uparrow)$ and $\sigma(\downarrow)$ are the electrical conductivity from spin-up and spin-down channel, respectively. For the other three alloys, two current model is not applicable. For all the member of this series, the calculated values of $S$ are found to be negative. The calculated $S$ for $Fe_2CrAl$, $Fe_{1.75}Mn_{0.25}CrAl$, $Fe_{1.5}Mn_{0.5}CrAl$, $Fe_{1.25}Mn_{0.75}CrAl$ and $FeMnCrAl$ are -6.92, -9.07, -5.75, -4.06 and -1.55 μV/K, respectively. The calculated and experimental are found to be consistent and exhibit good agreement with each other at 300 K. Also, in some temperature regimes, $S$ varies linearly with temperature (as shown in Figure 10). The regimes for which experimental and calculated data matches are 300-450 K ($Fe_2CrAl$, $Fe_{1.75}Mn_{0.25}CrAl$), 300-500 K ($Fe_{1.5}Mn_{0.5}CrAl$) 300-400 K ($Fe_{1.25}Mn_{0.75}CrAl$ and $FeMnCrAl$). The linear variation indicates that in this regime, the major contributions in $S$ are from the electrons. The observed deviation from the linear behaviour is ascribed to the dominance of bipolar effect. Due to the limitations of the BoltzTrap code, variation of $S$, above the deviation temperature, cannot be explained. However, by considering temperature dependent effect on electronic structure, scattering time parameters and with suitable theoretical model; it can be investigated.

4. Conclusion

$Fe_2CrAl$ is a thermoelectric material, with a of small band gap ~ 0.318 eV in spin-down channel. A slight substitution (12.5%) of Mn reduces this band gap because of the weak coupling between *d*-states of Fe/Mn and Cr atoms. Enhancement of this coupling leads to complete disappearance of band gap in higher Mn-substituted alloys. As compared to $Fe_2CrAl$, even though *S* and *ρ* are enhanced in $Fe_{1.75}Mn_{0.25}CrAl$, these parameters systematically decrease with an increase in Mn-substitution, due to absence of band gap and large *n*. However, across the series $\kappa_L$ is unaffected because of the absence of mass and volume fluctuations due to the comparable radii of Fe and Mn. Our studies reveal *S* and *ρ* shows a strong dependence on *n* and strength of *d-d* hybridization between Fe/Mn and Cr atoms. Our studies will be helpful for the researchers to understand the factors influencing the high temperature thermoelectric properties of full-Heusler alloys.

**Acknowledgements**

KM acknowledges financial support from DST-SERB project EMR/2016/00682.

**Disclosure statement**

No potential conflict of interest was reported by the authors.

**Table 1** Carrier concentration ($n$) of $Fe_{2-x}Mn_xCrAl$ at 300 K.

| Alloy | Carrier concentration, $n$ (per $cm^3$) |
|---|---|
| $Fe_2CrAl$ | $2.16 \times 10^{21}$ |
| $Fe_{1.75}Mn_{0.25}CrAl$ | $9.8 \times 10^{20}$ |
| $Fe_{1.5}Mn_{0.5}CrAl$ | $1.42 \times 10^{21}$ |
| $Fe_{1.25}Mn_{0.75}CrAl$ | $1.84 \times 10^{21}$ |
| $FeMnCrAl$ | $2.31 \times 10^{21}$ |

**Figures**

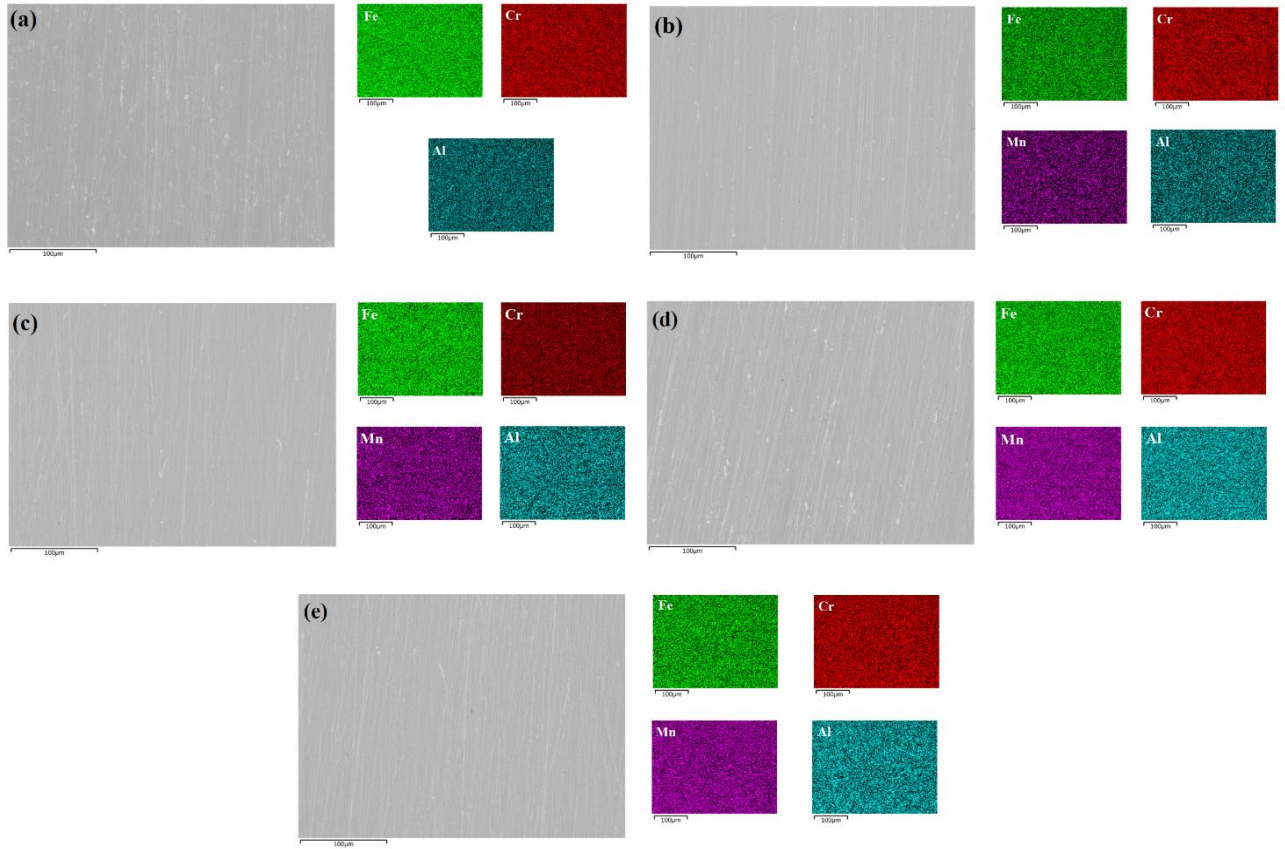

Figure 1. (a)-(e) represents (in the scale of 100 µm) the SEM images of $Fe_2CrAl$, $Fe_{1.75}Mn_{0.25}CrAl$, $Fe_{1.5}Mn_{0.5}CrAl$, $Fe_{1.25}Mn_{0.75}CrAl$ and FeMnCrAl, respectively. Corresponding EDS elemental mapping for Fe, Cr, Mn and Al in respective alloys are also shown.

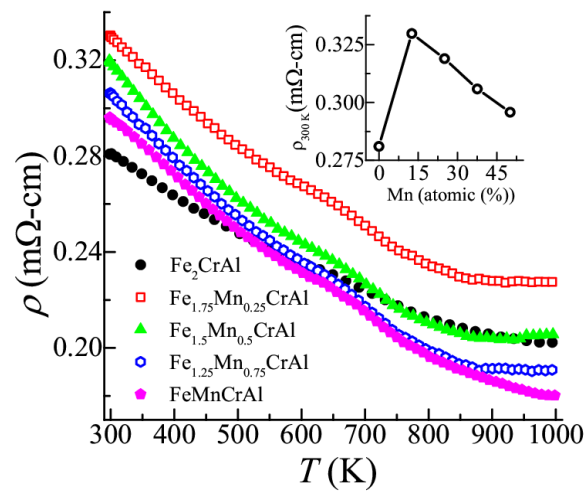

Figure 2. Electrical resistivity ($\rho$) of $Fe_{2-x}Mn_xCrAl$ (0≤x≤1) as a function of temperature in the range 300 – 950 K. Inset: $\rho$ value at 300 K as a function of Mn (atomic %) content.

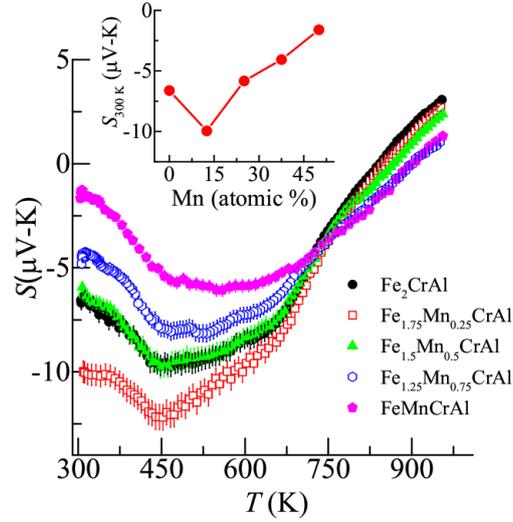

Figure 3. Temperature dependent behaviour of Seebeck coefficient (*S*) of all alloys in the range 300-950 K. Inset: *S* value at 300 K as a function of Mn (atomic %) content.

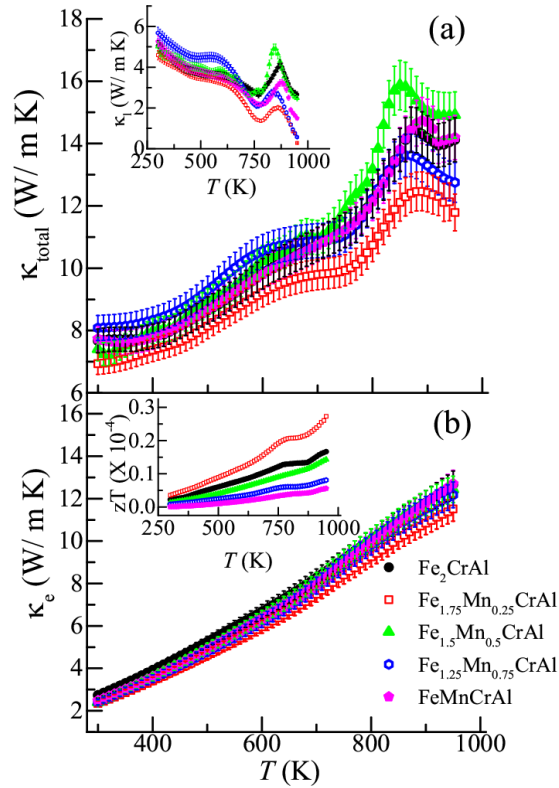

Figure 4. Temperature response of (a) total thermal conductivity ($\kappa_{total}$) (b) electronic thermal conductivity ($\kappa_e$) of all alloys in the range 300-950 K. Insets: Temperature response of (a) $\kappa_L$ and (b) zT of all alloys in the range 300-950K

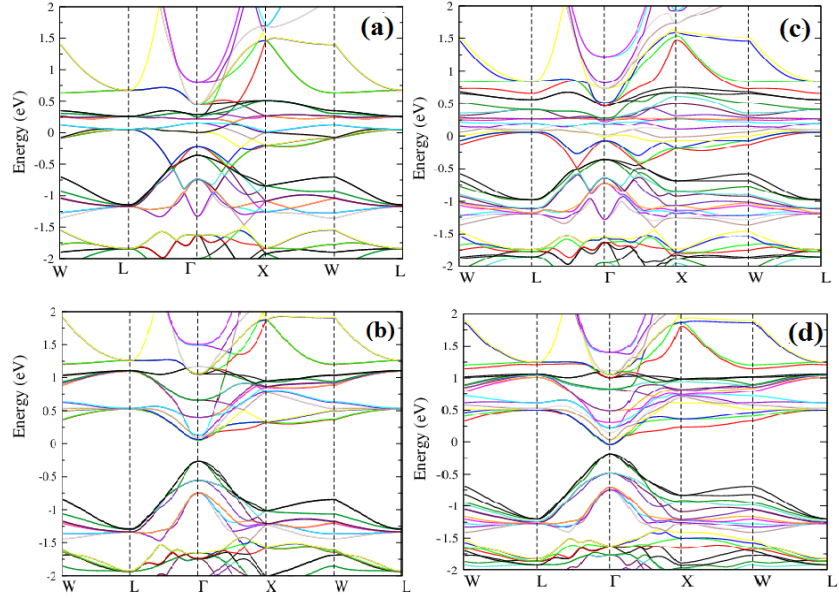

Figure 5 Spin polarized electronic band structures (a) spin-up (b) spin-down channel for Fe$_2$CrAl (c) spin-up (d) spin-down channel for Fe$_{1.75}$Mn$_{0.25}$CrAl.

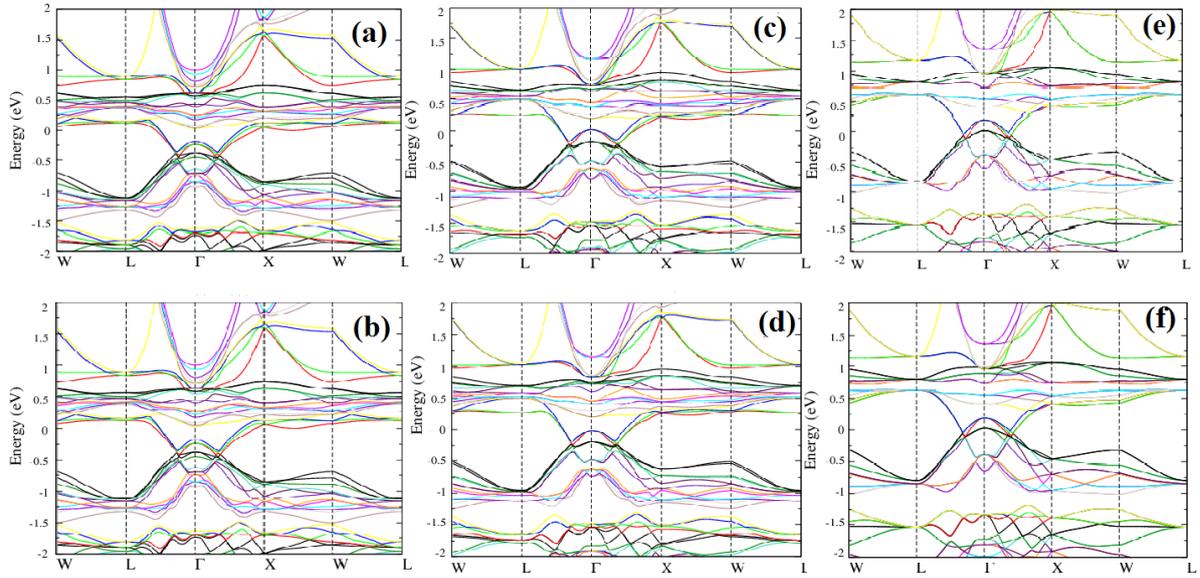

Figure 6. Spin polarized electronic band structures (a) spin-up (b) spin-down channel for Fe$_{1.5}$Mn$_{0.5}$CrAl; (c) spin-up (d) spin-down channel for Fe$_{1.25}$Mn$_{0.75}$CrAl; (e) spin-up (f) spin-down channel for FeMnCrAl

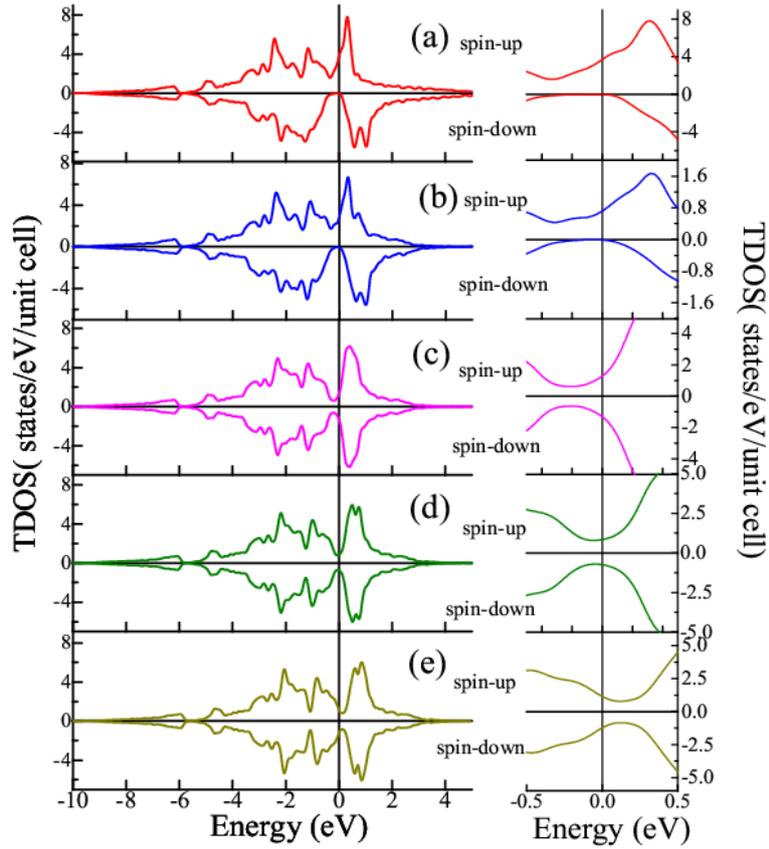

Figure 7. Left panel: Total Density of states (TDOS) plots of (a) $Fe_2CrAl$ (b) $Fe_{1.75}Mn_{0.25}CrAl$ (c) $Fe_{1.5}Mn_{0.5}CrAl$ (d) $Fe_{1.25}Mn_{0.75}CrAl$ (e) FeMnCrAl, respectively. Right panel: Same plot in the -0.5 to 0.5 eV energy range.

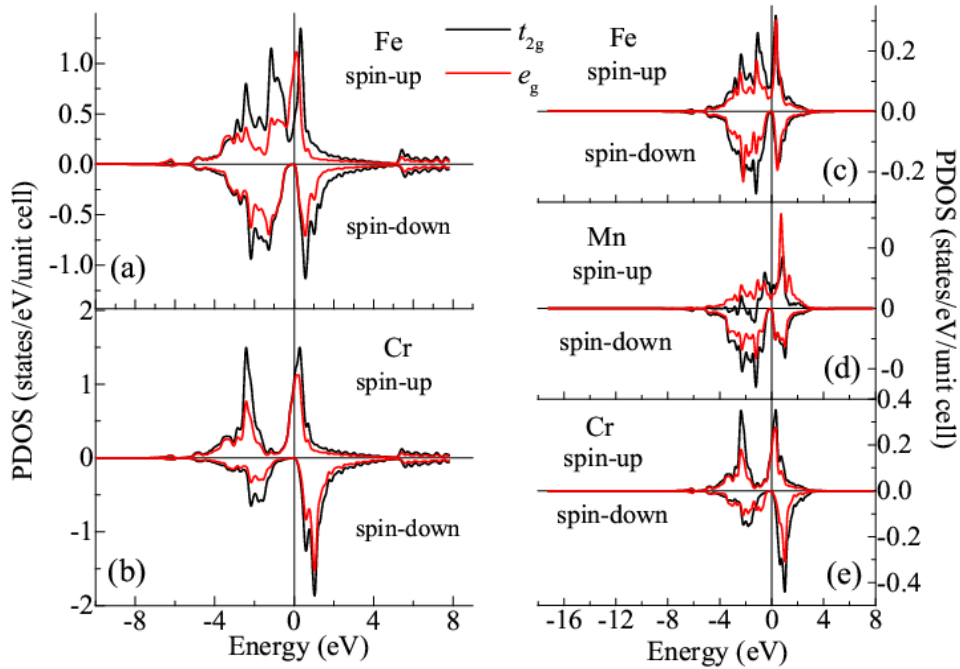

Figure 8. Partial Density of states (PDOS) plots of (a) Fe (b) Cr of $Fe_2CrAl$ ; (c) Fe (d) Mn (e) Cr of $Fe_{1.75}Mn_{0.25}CrAl$, respectively

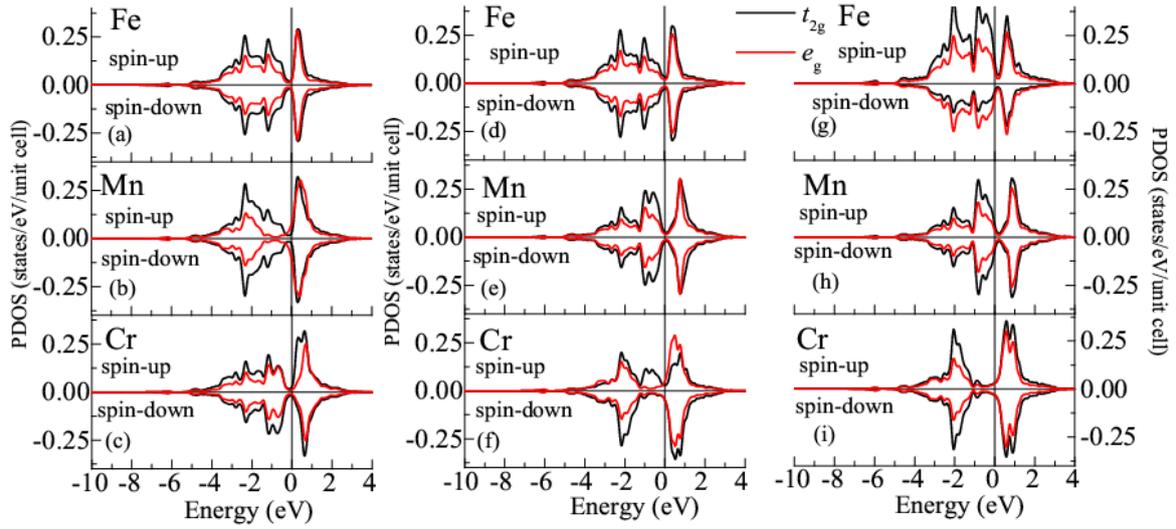

Figure 9. Partial Density of states (PDOS) plots of (a) Fe (b) Mn and (c) Cr of $Fe_{1.5}Mn_{0.5}CrAl$; (d) Fe (e) Mn (f) Cr of $Fe_{1.25}Mn_{0.75}CrAl$ and (g) Fe (h) Mn and (i) Cr of FeMnCrAl

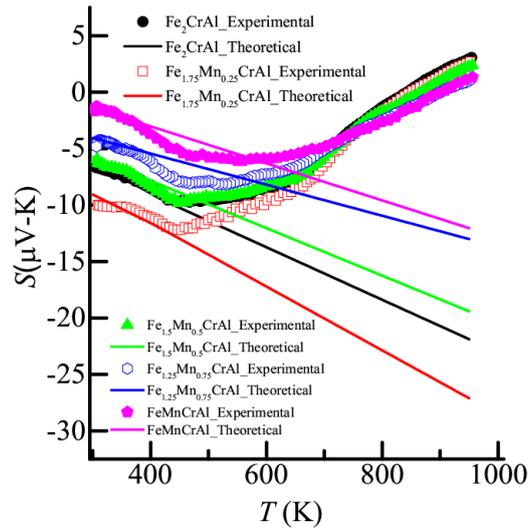

Figure 10. Comparison between experimental and theoretical temperature dependence of Seebeck coefficient (*S*) of all alloys in the temperature range 300-950 K.

Supplementary material for

# Anomalous dependence of thermoelectric parameters on carrier concentration and electronic structure in Mn-substituted $Fe_2CrAl$ Heusler alloy

Kavita Yadav[1], Saurabh Singh[2], Omprakash Muthuswamy[2], Tsunehiro Takeuchi[2], and K. Mukherjee[1]

[1] School of Basic Sciences, Indian Institute of Technology, Mandi, Himachal Pradesh-175005, India

[2] Research Centre for Smart Energy Technology, Toyota Technological Institute, Nagoya, 468-8511, Japan

**Supplementary Tables S1-S5** shows the typical chemical composition of $Fe_2CrAl$, $Fe_{1.75}Mn_{0.25}CrAl$, $Fe_{1.5}Mn_{0.5}CrAl$, $Fe_{1.25}Mn_{0.75}CrAl$ and $FeMnCrAl$ alloys by electron Probe Micro-Analyzer (EPMA) with nominal compositions. Four different crystal detectors (LIF, PET, LIFH, PETH) are used to collect EPMA spectra. The conditions set for the measurements of each point are under the of 20 kV acceleration voltage together with 20 µm of beam spot size. The measured compositions are calculated from the 9 randomly selected point scans of all the alloys.

**Supplementary Table S1**

| Nominal | $Fe_2CrAl$ | | |
|---|---|---|---|
| Point No. | Fe composition (%) | Cr composition (%) | Al composition (%) |
| 1 | 50.3271 | 27.1991 | 22.4739 |
| 2 | 50.1294 | 27.48 | 22.3906 |
| 3 | 50.1671 | 27.4702 | 22.3627 |
| 4 | 50.2556 | 27.2818 | 22.4626 |
| 5 | 50.0421 | 27.3832 | 22.5747 |
| 6 | 50.3055 | 27.4631 | 22.2315 |
| 7 | 50.566 | 27.2404 | 22.1936 |
| 8 | 50.4118 | 27.2941 | 22.2942 |
| 9 | 50.3578 | 27.3477 | 22.2945 |
| Average | **50.285** | **27.351** | **22.364** |
| St. Dev. | **0.158** | **0.105** | **0.124** |
| $Fe_{2.011\pm0.006}Cr_{1.094\pm0.004}Al_{0.8945\pm0.005}$ | | | |

**Supplementary Table S2**

| Nominal | Fe$_{1.75}$Mn$_{0.25}$CrAl | | | |
|---|---|---|---|---|
| Point No. | Fe composition (%) | Mn composition (%) | Cr composition (%) | Al composition (%) |
| 1 | 44.5423 | 6.1472 | 27.0225 | 22.288 |
| 2 | 44.4257 | 6.1502 | 27.119 | 22.3122 |
| 3 | 44.6915 | 6.1255 | 26.8774 | 22.3056 |
| 4 | 44.4746 | 6.1311 | 27.09 | 22.3043 |
| 5 | 44.5557 | 6.1113 | 27.1056 | 22.2274 |
| 6 | 44.5158 | 6.0995 | 27.0241 | 22.3606 |
| 7 | 44.4666 | 6.086 | 27.0958 | 22.3516 |
| 8 | 44.6738 | 6.099 | 26.9826 | 22.2446 |
| 9 | 44.5661 | 6.1374 | 27.0888 | 22.2077 |
| Aver. | **44.546** | **6.121** | **27.045** | **22.289** |
| St. Dev. | **0.090** | **0.023** | **0.078** | **0.053** |
| Fe$_{1.782\pm0.004}$Mn$_{0.245\pm0.001}$Cr$_{1.082\pm0.003}$Al$_{0.8916\pm0.002}$ | | | | |

**Supplementary Table S3**

| Nominal | Fe$_{1.50}$Mn$_{0.50}$CrAl | | | |
|---|---|---|---|---|
| Point No. | Fe composition (%) | Mn composition (%) | Cr composition (%) | Al composition (%) |
| 1 | 38.5591 | 12.5575 | 25.0156 | 23.8978 |
| 2 | 38.4591 | 12.5647 | 25.0894 | 23.8339 |
| 3 | 38.5365 | 12.6051 | 25.0068 | 23.8516 |
| 4 | 38.585 | 12.5928 | 25.1136 | 23.7086 |
| 5 | 38.7262 | 12.5702 | 25.0595 | 23.6442 |
| 6 | 38.5924 | 12.6155 | 25.1019 | 23.6902 |
| 7 | 38.608 | 12.551 | 25.003 | 23.8339 |
| 8 | 38.6185 | 12.7742 | 25.2497 | 23.3575 |
| 9 | 38.604 | 12.561 | 25.096 | 23.739 |
| Aver. | 38.588 | 12.600 | 25.082 | 23.729 |
| Std. Dev. | 0.071 | 0.069 | 0.076 | 0.163 |
| Fe$_{1.543\pm0.003}$Mn$_{0.504\pm0.003}$Cr$_{1.003\pm0.003}$Al$_{0.949\pm0.007}$ | | | | |

**Supplementary Table S4**

| Nominal | Fe$_{1.25}$Mn$_{0.75}$CrAl | | | |
|---|---|---|---|---|
| Point No. | Fe composition (%) | Mn composition (%) | Cr composition (%) | Al composition (%) |
| 1 | 32.0421 | 17.8092 | 26.7669 | 23.3819 |
| 2 | 32.0369 | 17.8613 | 26.8409 | 23.2609 |
| 3 | 32.7833 | 17.4999 | 27.3245 | 22.3922 |
| 4 | 32.1747 | 17.8031 | 26.6455 | 23.3767 |
| 5 | 32.1304 | 17.851 | 26.9728 | 23.0458 |
| 6 | 32.7681 | 17.2482 | 27.7403 | 22.2434 |
| 7 | 32.1035 | 17.7929 | 26.7866 | 23.317 |
| 8 | 32.0622 | 17.8098 | 27.0629 | 23.0651 |
| 9 | 32.1344 | 17.8347 | 26.8029 | 23.2279 |
| Aver. | **32.248** | **17.723** | **26.993** | **23.035** |
| St. Dev. | **0.302** | **0.209** | **0.344** | **0.425** |
| | Fe$_{1.289\pm0.012}$Mn$_{0.710\pm0.008}$Cr$_{1.079\pm0.013}$Al$_{0.921\pm0.017}$ | | | |

**Supplementary Table S5**

| Nominal | FeMnCrAl | | | |
|---|---|---|---|---|
| Point No. | Fe composition (%) | Mn composition (%) | Cr composition (%) | Al composition (%) |
| 1 | 27.3966 | 22.016 | 27.8245 | 22.763 |
| 2 | 27.2292 | 22.5929 | 27.5042 | 22.6738 |
| 3 | 27.4413 | 22.4614 | 27.5007 | 22.5966 |
| 4 | 27.4 | 22.0387 | 28.0681 | 22.4932 |
| 5 | 27.5798 | 22.2128 | 27.7929 | 22.4145 |
| 6 | 27.2708 | 22.3789 | 27.7889 | 22.5614 |
| 7 | 27.2054 | 22.3256 | 27.9426 | 22.5264 |
| 8 | 27.3804 | 22.0121 | 28.2089 | 22.3986 |
| 9 | 27.2283 | 22.4687 | 27.6884 | 22.6146 |
| Aver. | **27.348** | **22.279** | **27.813** | **22.56** |
| St. Dev. | **0.124** | **0.219** | **0.237** | **0.118** |
| | Fe$_{1.094\pm0.004}$Mn$_{0.891\pm0.009}$Cr$_{1.113\pm0.009}$Al$_{0.902\pm0.005}$ | | | |

**Supplementary Table S6-** The theoretical mass density of the $Fe_{2-x}Mn_xCrAl$ ($0 \leq x \leq 1$)

| Alloys | Theoretical density (g/cc$^3$) |
|---|---|
| $Fe_2CrAl$ | 6.540 |
| $Fe_{1.75}Mn_{0.25}CrAl$ | 6.512 |
| $Fe_{1.5}Mn_{0.5}CrAl$ | 6.474 |
| $Fe_{1.25}Mn_{0.75}CrAl$ | 6.453 |
| $FeMnCrAl$ | 6.410 |

**Supplementary Table S7-** Theoretical (from DFT calculations within GGA approximation) and experimental lattice parameters (Rietveld Refinement) of the $Fe_{2-x}Mn_xCrAl$ ($0 \leq x \leq 1$)

| Alloys | Experimental lattice parameter $a$ (Å) [Ref. 24] | Theoretical lattice parameter $a$ (Å) |
|---|---|---|
| $Fe_2CrAl$ | 5.784 | 5.652 |
| $Fe_{1.75}Mn_{0.25}CrAl$ | 5.791 | 5.658 |
| $Fe_{1.5}Mn_{0.5}CrAl$ | 5.800 | 5.664 |
| $Fe_{1.25}Mn_{0.75}CrAl$ | 5.804 | 5.675 |
| $FeMnCrAl$ | 5.815 | 5.683 |

**Effect of anti-site disorder (between Fe and Al) on the electronic band structure of $Fe_{2-x}Mn_xCrAl$ ($0 \leq x \leq 1$) using KKR-CPA calculations**

**Computational details:**

In order to see the effect of Fe-Al antisite disorder to the electronic structure of $Fe_{2-x}Mn_xCrAl$ (x = 0.25, 0.5, 0.75, 1.0) systems, we have performed the calculations of self-consistency field (scf) and estimated the density of states using charge self-consistent Korringa-Kohn-Rostoker (KKR) method implemented in the Akai-KKR program package. In this program, the anti-site disorder effect is employed within the coherent potential approximation (CPA). The crystal potential of muffin-tin form was established within the Generalized Gradient Approximation. The PBE exchange-correlation functional were taken into account for all the calculations. High dense k-points in the Brillouin zone of size equal to 14, and energy convergence criteria for total energy were set to 0.0001 eV. For all the compositions the magnetic configuration with scaler relativistic approximation was chosen. For the estimation of density of state plot tetrahedral integration method was used. The experimental lattice parameters corresponding to each composition obtained from the Rietveld refinement in the earlier work were used for calculation. The anti-site disorder incorporated between Fe and Al site from 0 to 10%. In our discussions we have used x% Fe-Al anti-site disorder to depict the exchange of x% of the Fe atom number at Al site and x% of Al atom number at Fe site.

**Results and Discussions:**

The total density of states (TDOS) of each alloy corresponding to 0%, 4% and 10 % anti-site disorder between Fe and Al has been plotted and shown in Figure S1. It has been noted that $Fe_2CrAl$ and $Fe_{1.75}Mn_{0.25}CrAl$ exhibit characteristics similar to half metallic ferromagnets. This observation is analogy with results obtained through WEIN2K code. The value of band gaps obtained corresponding to $Fe_2CrAl$ and $Fe_{1.75}Mn_{0.25}CrAl$ composition has been given in supplementary table S8. For $Fe_2CrAl$, the value of band gap (in spin down channel) has been found ~ 0.24 eV, which is lower than that obtained using WEIN2K. It has been noted that the value of band gap reduces with increment in Fe-Al anti-site disorder. In $Fe_{1.75}Mn_{0.25}CrAl$, the band gap is determined to be ~ 0.21 eV, which is lower than $Fe_2CrAl$. It is smaller than the band gap obtained using WEIN2K code. This discrepancy reflects that KKR-CPA method is more accurate to determine the half metallicity in the material. In $Fe_2CrAl$, it can be seen that there is drastic decrement in total density of states (TDOS) in the vicinity of fermi level ($E_F$) with increment in anti-site disorder. This reflects that the resistivity of the parent alloy will increase on enhancement in anti-site disorder. However, in $Fe_{1.75}Mn_{0.25}CrAl$, a small change in TDOS is noted with increment in anti-site disorder. This indicates that there will be negligible effect on the magnitude of resistivity. Whereas, in other alloys, metallic behaviour is noted in spin down channel. Additionally, it has been observed that with increasing concentration (x<0.25), the TDOS in the vicinity of fermi-level increases. Moreover, with Mn concentration (x<0.25), there is systematic shift in TDOS at $E_F$ towards higher energy. This is in analogy with the observed trend in magnitude of $\rho$ at 300 K, in the latter three compositions. It can be concluded that presence of anti-site disorder in $Fe_2CrAl$ and $Fe_{1.75}Mn_{0.25}CrAl$, adversely affect the magnitude of band gap, whereas, in other Mn doped alloys, it has negligible effect.

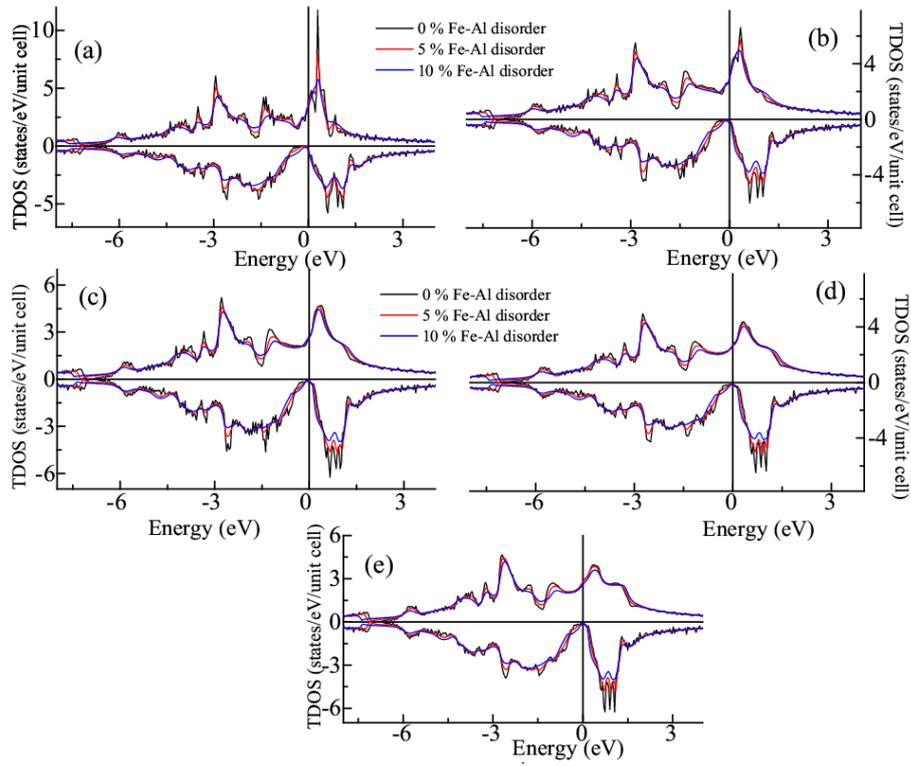

**Figure S1** Total Density of states (TDOS) plots of (a) $Fe_2CrAl$ (b) $Fe_{1.75}Mn_{0.25}CrAl$ (c) $Fe_{1.5}Mn_{0.5}CrAl$ (d) $Fe_{1.25}Mn_{0.75}CrAl$ (e) $FeMnCrAl$, respectively.

**Supplementary Table S8-** The value of band gaps of $Fe_2CrAl$ and $Fe_{1.75}Mn_{0.25}CrAl$ determined from the electronic band structure corresponding to 0 %, 4% and 10 % anti-site disorder.

| Band gaps | $Fe_2CrAl$ | $Fe_{1.75}Mn_{0.25}CrAl$ |
|---|---|---|
| 0% anti-site disorder | ~0.24 eV | ~0.21 eV |
| 4 % anti-site disorder | ~0.19 eV | ~0.17 eV |
| 10% anti-site disorder | ~0.14 eV | ~0.10 eV |